\documentclass[12pt]{article}
\usepackage{titlesec}
\titleformat{\paragraph}{\normalfont\normalsize\bfseries}{\theparagraph}{1em}{}
\titlespacing*{\paragraph}{0pt}{3ex plus 1.5ex minus .5ex}{1.5ex plus .5ex}

\usepackage{amsmath}
\usepackage{amssymb}

\usepackage{graphicx}

\usepackage{cite}
\usepackage{soul}
\usepackage{color} 
\usepackage{authblk}
\usepackage{xfrac}
\usepackage{bm}
\usepackage{geometry}
\usepackage{graphicx}
\usepackage{epstopdf}
\graphicspath{{./figures/}}
\DeclareGraphicsExtensions{.eps}
\usepackage{setspace} 
\usepackage[labelfont=bf,labelsep=period,justification=raggedright]{caption}

\date{}

\title{The Radical Pair Mechanism and the Avian Chemical Compass: Quantum Coherence and Entanglement}
\author[1]{Yiteng Zhang}
\affil[1]{Department of Physics and Astronomy, Purdue University, West Lafayette, IN, 47907 USA}
\author[2]{Gennady P. Berman}
\affil[2]{Theoretical Division, LANL, and New Mexico Consortium, Los Alamos, NM 87545 USA}
\author[3,4]{Sabre Kais\thanks{kais@purdue.edu}}
\affil[3]{Department of Chemistry, Department of Physics and Astronomy and Birck Nanotechnology Center, Purdue University, West Lafayette, IN 47907 USA}
\affil[4]{Qatar Environment and Energy Research Institute, Qatar Foundation, Doha, Qatar}

\begin{document}

\maketitle
\section*{Abstract}
We review the spin radical pair mechanism which is a promising explanation of avian navigation. This mechanism is based on the dependence of product yields on 1) the  hyperfine interaction involving electron spins and  neighboring nuclear spins and 2) the intensity and orientation of the geomagnetic field. This review describes the general scheme of chemical reactions involving radical pairs generated from singlet and triplet precursors; the spin dynamics of the radical pairs; and the magnetic field dependence of product yields caused by the radical pair mechanism. The main part of the review includes a description of the chemical compass in birds. We review: the general properties of the avian compass; the basic scheme of the radical pair mechanism; the reaction kinetics in cryptochrome; quantum coherence and entanglement in the avian compass; and the effects of noise. We believe that the quantum avian compass can play an important role in avian navigation and can also provide the foundation for a new generation of sensitive and selective magnetic-sensing nano-devices. 
\newpage

\section{Radical Pair Mechanism}

	\subsection{Introduction}
A radical is an atom, molecule or ion that has unpaired valence electrons. Radicals and radical pairs often play a very important role as intermediates in thermal, radiation, and photochemical reactions \cite{dyn1}. The presence of unpaired electron spins in these systems allows one to influence and control these reactions using interactions between external magnetic fields and electron spins \cite{dyn2}. However, until 1970, most scientists believed that ordinary magnetic fields had no significant effect on chemical or biochemical reactions, since the magnetic energy of typical molecules, under ordinary magnetic fields, is much smaller than the thermal energy at room temperature and is much smaller than the activation energies for those reactions \cite{dyn1,dyn2}. This situation changed significantly in the 1970's after a series of experimental results were reported on magnetic field effects on chemical reactions \cite{exp1,exp2,exp3,exp4,exp5}. Because of these experimental studies, a number of researchers have made an effort to theoretically explain the magnetic field effects on the chemical reactions\cite{theo1,theo2}. Thanks to these and the subsequent efforts, we are now able to explain systematically magnetic field effects in terms of the radical pair mechanism. The radical pair mechanism  was then successfully applied to explain the chemically induced nuclear polarization and electron polarization, which were shown to be based on the spin dynamics of radical pairs \cite{dyn2}. 
\begin{figure}[htbp]
\includegraphics[width=0.8\textwidth]{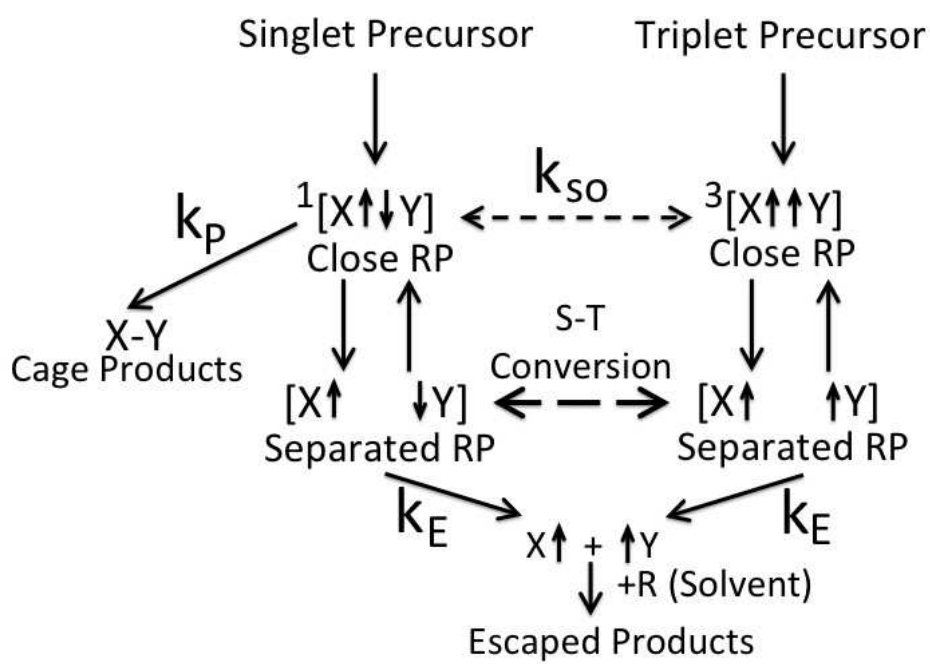}
\caption
{\small{Reaction scheme of radical pairs generated from singlet and triplet precursors. Singlet and triplet radical pairs are represented by $^{1}\left[...\hspace{1mm} \right]$ and 
		$^{3}\left[...\hspace{1mm} \right]$, respectively.  $k_{SO,P,E}$ are the rates of reactions. (Taken from Ref.\cite{dyn2}, with modifications.)}}
\label{RPsch}
\end{figure}
\\
\\
According to the radical pair mechanism, an external magnetic field affects chemical reactions by alternating the electron spin state of a weakly coupled radical pair, which is produced as an intermediate. The basic scheme of chemical reactions through the radical pairs is shown in Fig. \ref{RPsch}. Radical pairs are usually produced as short-lived intermediates through decomposition, electron transfer, or hydrogen transfer reactions from singlet or triplet excited states. These reaction precursors are called ``S-precursors" or ``T-precursors". 
\\
\\
The generated radical pairs are surrounded by a solvent molecular cluster, called a ``solvent cage", and these pairs retain the spin multiplicity of their precursors. Initially, two radicals are close together, and are called  the ``close pair". Sometimes, recombination reactions occur from S- or T-close pairs immediately after the formation of the radical pairs. Such reactions are called ``primary recombinations",  and their products are called ``cage products". However, because of the Pauli exclusion principle, the T-close pairs require enough energy to produce excited ``cage products", while the S-close pairs are able to produce the ground state of the ``cage products", due to the spin preservation during the chemical reactions. Consequently, the recombination reactions from T-close pairs occur less frequently than those from S-close pairs. Usually, we can ignore reactions from T-close pairs. However, the singlet-triplet conversion is possible for close pairs involving heavy atom-centered radicals due to their spin-orbit interactions. But for close pairs involving only light atom-centered radicals, no spin conversion occurs between their singlet and triplet states. The S-T conversion mainly occurs during the second stage. In the second stage, as shown in Fig. \ref{RPsch}, the two radicals begin to diffuse away from each other, forming a separate pair. When the two radicals are separated at a certain distance, the S-T conversion becomes possible through weak magnetic interactions of radicals including Zeeman and the hyperfine interactions, as will be explained in detail later. In the last stage, some of the separated radicals approach each other, forming close pairs again, and some continue to diffuse from each other, forming free radicals and producing ``escape products" with or without the solvent molecules \cite{dyn1,dyn2}. 
	\subsection{Spin Dynamics of the Radical Pairs}
Consider two weakly coupled radicals that form a radical pair. The spin dynamics of the radical pair is governed by a Hamiltonian ($\vec{H}_{\small{RP}}$), which can be expressed as the sum of an exchange term ($\vec{H}_{\small{ex}}$) and a magnetic ($\vec{H}_{\small{mag}}$) term \cite{dyn2,theo1},
\begin{equation}
\vec{H}_{\small{RP}} = \vec{H}_{\small{ex}} + \vec{H}_{\small{mag}} \label{Hrp},
\end{equation}
where, 
\begin{equation}
\vec{H}_{\small{ex}} = -J(r)\biggl(2\vec{S}_{1}\cdot \vec{S}_{2}+\frac{1}{2}\biggr)\label{Hex},
\end{equation}
\begin{equation}
\vec{H}_{\small{mag}} = \mu_B\vec{B} \cdot (g_1\vec{S}_{1}+g_2\vec{S}_{2})+(\sum_i^a{A}_{1i}\vec{S}_1 \cdot \vec{I}_{1i}+\sum_k^b{A}_{2k}\vec{S}_2 \cdot \vec{I}_{2k}) \label{Hmag}.
\end{equation}
In Eq. \ref{Hex}, $\vec{S}_i=\frac{1}{2}\vec{\sigma}_i$, where $i = 1, 2$, and $\vec{\sigma}_i$ are the Pauli matrices, $J(r)$ is the value of the exchange integral between two unpaired electron spins ($\vec{S}_1$ and $\vec{S}_2$), which decreases with separation distance, $r$. In Eq. \ref{Hmag}, the first two terms describe the Zeeman effects, and the last two terms are hyperfine interactions between the electron spins ($\vec{S}_1$, $\vec{S}_2$) and the nuclear spins ($\vec{I}_{1i}$, $\vec{I}_{2k}$) in the radicals 1 and 2. Nuclear Zeeman effects are neglected since their magnitudes are much smaller than those of the electron Zeeman terms and hyperfine coupling terms. Also, $g_1$ and $g_2$ are the isotropic $g$-values of the two component radicals in the radical pair, respectively, and ${A}_{1i}$ and ${A}_{2k}$ are the isotropic hyperfine coupling constants in radicals 1 and 2, respectively, and the number of nuclei in radicals 1 and 2 are $a$ and $b$, respectively. 
\\
\\
\begin{figure} [htbp]
\begin{center}
\includegraphics[width=0.75\textwidth]{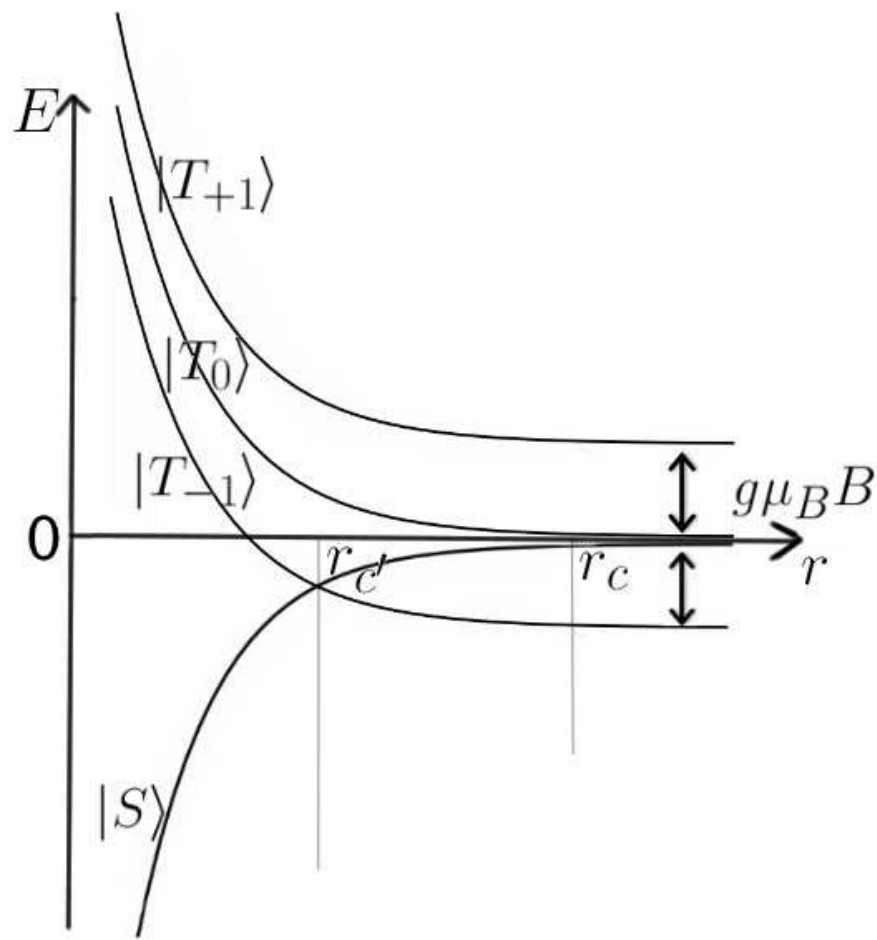}
\end{center}
\caption{\small{Dependence of a radical pair's energy on the distance ($r$) between two components when $B\ne 0$ and $J(r)$ is negative.} (Taken from Ref.\cite{dyn1}, with modifications.)} 
\label{energy}
\end{figure}
The state of a radical pair can be represented by the product of the electron and nuclear states. The two unpaired electron spins generate the singlet ($|S\rangle$) and triplet ($|T_n\rangle$; $n$ = -1, 0, 1) states which are expressed as:
\begin{eqnarray}
|S\rangle & = & \frac{1}{\sqrt{2}}(\mid\uparrow\downarrow\rangle-\mid\downarrow\uparrow\rangle)),  \\
|T_0\rangle & = & \frac{1}{\sqrt{2}}(\mid\uparrow\downarrow\rangle+\mid\downarrow\uparrow\rangle)), \\
|T_{+1}\rangle & = & \mid\uparrow\uparrow\rangle, \\
|T_{-1}\rangle & = & \mid\downarrow\downarrow\rangle. 
\end{eqnarray}
We use $\mid\uparrow\rangle$ or $\mid\downarrow\rangle$ to express the $z$-component of a electron spin state, and we  use $|M \rangle$ to express the $z$-component of a nuclear spin state, i.e. $\hat{I}_z|M\rangle = M|M\rangle$. Since we ignore the interactions between nuclear spins, the total nuclear spin states can be expressed as, $|\chi_N\rangle = \prod_i^a|M_i\rangle\prod_k^b|M_k\rangle$, where $a$ and $b$ are the number of nuclei in radicals 1 and 2, respectively. 
\\
\\
In the presence of an external magnetic field, we can calculate the energies of the singlet and the triplet radical pairs,
\begin{equation}
E(S) = \langle S, \chi_N|\vec{H}_{RP}|S, \chi_N\rangle = J(r),
\end{equation}
\begin{equation}
E(T_n) = \langle T_n, \chi_N|\vec{H}_{RP}|T_n, \chi_N\rangle= -J(r) + ng\mu_BB+\frac{n}{2}(\sum_i^a {A}_{1i} M_i + \sum_k^b{A}_{2k}M_k),
\end{equation}
where,  $n=$ -1, 0 and +1; $g=\frac{g_1+g_2}{2}$.
\\
\\
The $r$-dependent energy of a radical pair in the external magnetic field is schematically depicted in Fig. \ref{energy} in the range of a few molecular diameters. Since the exchange interaction ($J(r)$) decreases with the distance exponentially, $J(r)$ can be neglected safely in the second stage ($r > r_c$) of radical pair mechanism mentioned before. Therefore, for further discussion, we will neglect the exchange interaction ($J(r)$). 
\\
\\
When $r > r_c$, the conversion between $|S\rangle$ and $|T_n\rangle$ is governed by the following rules:
\begin{equation}
\langle T_0 , \chi_N|\vec{H}_{RP}|S, \chi_N\rangle = \frac{1}{2}(\Delta g \mu_BB+(\sum_i^a{A}_{1i}M_i-\sum_k^b{A}_{2k}M_k)) = \hbar\omega_N, \label{ST0}
\end{equation}
\begin{equation}
\langle T_{\pm 1} , \chi_{N'}|\vec{H}_{RP}|S, \chi_N\rangle = \frac{\mp A_{1i}}{2\sqrt{2}}[I_{1i}(I_{1i}+1) - M_i(M_i\mp 1)]^{\frac{1}{2}} \label{ST1},
\end{equation}
where, $\Delta g=g_1-g_2$, $|\chi_N\rangle = |M_i, M_k\rangle$, $|\chi_{N'}\rangle = |M'_i, M'_k\rangle$, $M'_i = M_i \mp 1$, $M'_k=M_k$, and $\omega_N$ is the nuclear Rabi frequency.
\\
\\
 It is worth mentioning that in large external magnetic fields, due to the large energy gap ($g\mu_BB$), only the $S-T_0$ conversion is taken into account, and the $S-T_{\pm1}$ conversions can be neglected \cite{kaph}. However, in small magnetic fields, all conversions between singlet and three triplet states of the radical pair must be considered, since the Zeeman terms are small compared to hyperfine interaction. 
\\
\\
By solving the Schr{\"o}dinger equation (\ref{schrod}), Kaptein obtained the time evolution of the wave function ($\Phi(t)$) of a radical pair during the $S-T_0$ conversion\cite{kaph}, 

\begin{equation}
i\hbar\frac{\partial \Phi(t)}{\partial t} = H_{RP} \Phi(t) \label{schrod},
\end{equation}
\begin{equation}
\Phi(t) = C_{SN}(t)|S, \chi_N\rangle + C_{T_0N}|T_0, \chi_N\rangle \label{wave},
\end{equation}
where, $C_{SN}(t) = C_S(0)\cos\omega_N t - iC_{T_0}(0)\sin\omega_N t$, $C_{T_0N}(t) = C_{T_0}(0)\cos\omega_N t - iC_S(0)\sin\omega_N t$. For chemical reactions occurring from S-precursors, $|C_{SN}(0)|=1$ and $|C_{T_0N}(0)| = 0$. In these cases, the wave function of a radical pair is, $\Phi(t) = (\cos\omega_N t)|S, \chi_N\rangle + (\sin\omega_N t)|T_0, \chi_N\rangle$. 
\\
\\
Ideally, the time evolution of the populations of singlet and triplet states are $\cos^2\omega_N t$ and $\sin^2\omega_N t$ during the $S-T_0$ conversion, respectively. However, due to random-walk diffusion within the solvent cages, the radicals either separate from each other further, forming ``escape radicals", or re-encounter with each other, realizing a secondary recombination. Therefore, the solvent can affect the lifetime of the generated radical pair. The more viscous the solvent is, the longer the lifetime of the radical pair becomes.

\subsection{Magnetic Field Effects on Product Yields due to the Radical Pair Mechanism}
The magnetic field can have an influence on the products of chemical reactions through the radical pairs in solution \cite{kapl}. Since the lifetime of a radical pair becomes longer in a more viscous solvent, the magnetic field effects, due the radical pair mechanism, become appreciable if the solvent is very viscous \cite{exp2}.
\\
\\
Eq. (\ref{ST0}) and Eq. (\ref{ST1}) clearly show that the $S-T$ conversion of a radical pair can be influenced by the external magnetic field and the hyperfine interaction, in the region where the exchange interaction ($J(r)$) can be neglected. Thus, the magnetic field effects on radical pairs in this region can be classified by two typical mechanisms: the $\Delta g$ mechanism and the hyperfine coupling mechanism. The $\Delta g$ mechanism is appreciable when $\Delta g \ne 0$ and $A_{1i} = A_{2k} = 0$, while the hyperfine coupling mechanism is appreciable when $\Delta g = 0$, and $A_{1i} \ne 0$ or $A_{2k} \ne 0$ \cite{dyn1,dyn2}.
\\
\\
For the $\Delta g$ mechanism, there is no $S-T$ conversion in the absence of external magnetic field. However, according to Eq. (\ref{ST0}), in an external magnetic field, the $S-T_0$ conversion can be induced with a frequency of $\sfrac{\Delta g\mu_BB}{2\hbar}$,  but the $S-T_{\pm 1}$ conversion will still not occur. This means that, due to the $\Delta g$ mechanism, the rate of $S-T$ conversion becomes larger with a stronger external magnetic field. In this situation, if the radical pair in a chemical reaction is produced from an S-precursor, the cage product will decrease as the magnetic field increases. Because a singlet radical pair is generated initially, the conversion rate will increase as the field increases. Also, it can be proved that the magnetically induced decrease in the yield of the cage product is proportional to $B^{\frac{1}{2}}$ \cite{dyn1}. On the other hand, the yield of products from the triplet radical pair, such as escape products, increases with  $B$ increasing. The opposite results occur when the radical pair is generated from a T-precursor. The yield of cage product increases, with $B$ increasing at a rate proportional to $B^{\frac{1}{2}}$. The yield of products from the triplet pair, such as escape products, decreases.
\\
\\
For the hyperfine coupling mechanism at zero magnetic field, the $S-T$ conversion can occur between the singlet state and all of the three sub-levels of the triplet state, through the hyperfine coupling terms in Eq. (\ref{ST0}) and Eq. (\ref{ST1}). In the presence of an external magnetic field, the three triplet sub-levels are split by the Zeeman effect. So the $S-T_{\pm 1}$ conversion will be forbidden due to the energy gap, but the $S-T_{0}$ conversion can occur with the same frequency, $\sfrac{(\sum_i^aA_{1i}M_i-\sum_k^bA_{2k}M_k)}{2\hbar}$, as the frequency without the magnetic field. As a result, the rate of the $S-T$ conversion decreases as the external magnetic field increases. And this decrease becomes saturated at higher fields. Due to this mechanism, if the radical pair is produced from the S-precursor, the yield of cage products will increase, and the yield of products produced from the triplet pair will decreases, when a non-zero magnetic field is present. The magnetically reduced triplet yield is characterized by the field, $B_{\sfrac{1}{2}}$, at a half saturation. Weller $et$ $al$. \cite{afh} gave an expression of the $B_{\sfrac{1}{2}}$ value as, $B_{\sfrac{1}{2}} = \frac{1}{2}(B_1+B_2)$, where $B_l = (\sum_jA^2_{lj}I_{ij}(I_{lj}+1))^{\frac{1}{2}}$ ($l = 1~\text{or}~2$; $j = 1,2,...,a$ for $l =1$ and $j = 1,2,...,b$ for $l = 2$). The opposite occurs for a pair generated from a T-precursor. The yield of cage product decreases with $B$ increasing at a rate proportional to $B^{\frac{1}{2}}$. The yield of products produced from the triplet pair, such as escape products, increases with $B$, and these magnetically induced changes should also become saturated at very high magnetic fields ($B\gg B_{\sfrac{1}{2}}$).
\\
\\
It is noteworthy that due to the $\Delta g$ mechanism and the hyperfine coupling mechanism, the external magnetic field has opposite effects on the product yields. The general case will be a mixture of these two mechanisms, meaning $\Delta g \ne 0$, and $A_{1i} \ne 0$ or $A_{2k} \ne 0$. In this case, the magnetic field effects, due to the $\Delta g$ mechanism and the hyperfine coupling mechanism, appear simultaneously. Since the magnetic field effects due to the hyperfine coupling mechanism have a saturated region, the magnetic field effects due to the $\Delta g$ mechanism dominate at high fields, and the magnetic field effects due to the hyperfine coupling mechanism dominate at low fields. Therefore, the cage product yield has a maximum (nadir), when the radical pair is generated from an S-precursor (T-precursor) as a function of increasing magnetic field.
\section{Chemical Compass in Birds}
\subsection{Introduction}
Before the twentieth century, biology and physics rarely crossed paths. But after the development of quantum mechanics, Schr{\"o}dinger introduced quantum physics into the realm of biology \cite{life}. However, quantum mechanics normally applies at the microscopic regions of the order of nanometers, and living organisms belong to the macroscopic world. Usually, quantum collective (coherent) effects can only be observed in laboratories using a high vacuum and/or an ultra-low temperature. Despite these differences, the progress in quantum biology has been very rapid since the twentieth century \cite{ps10, ac3,mfeck, ac6}. In some sense, the quantum mechanics always plays a significant role in biology since every chemical reaction relies on quantum mechanics, and chemical reactions are the basic processes in biological systems \cite{qb0, qb1,qb2,qb3, qb4}. Recently, many articles have been published in the field of quantum biology, including articles on photosynthesis \cite{ps1,ps2,ps3,ps4,ps5,ps6, ps7, ps8, ps9, ps10, ps11, ps12, ps13, ps14, ps15, ps16, ps17,ps18, ps19,ps20}, avian compass \cite{cc0,ac1,ac2,ac3,mfeck, ac4,ac5,ac6, ac7, ac8, ac9, bced, accb, vqb, ara}, and olfaction \cite{ol1,ol2}.
\\
\\
The precise mechanism of avian navigation has been a mystery for centuries. Every year migrant birds navigate hundreds or even thousands of miles between their seasonal habitats. Scientists have been trying to explain this astonishing phenomenon for decades. To date, the primary biophysical process underlying the avian compass still remains unknown. Scientists found that many sources of directional information can contribute to birds' navigation, such as landmarks or topographic features \cite{lm1,lm2}, stars \cite{star1, star2, star3, star4, star5}, the sun \cite{sun1, sun2, sun3}, and so on. Among these sources and hypotheses, magnetite as well as biochemical radical-pair reactions has been hypothesized to mediate sensitivity to Earth-strength magnetic fields through fundamentally different physical mechanisms \cite{res, psc}. In the magnetite-based mechanism, magnetic fields exert mechanical forces \cite{mt1, mt2}. Recently, Treiber $et$ $al$. reported that the iron minerals (located in cells in the upper beak of pigeons, and possibly magnetite), are actually macrophages and have nothing to do with magnetic sensing \cite{mt3, mcd}, but the magnetite-based mechanism is still a reasonable hypothesis to explain the magnetoreception in birds. In the radical-pair mechanism, magnetic fields alter the dynamics of transitions between spin states, after the creation of a radical pair through a light-induced electron transfer. These transitions in turn affect reaction rates and products \cite{cc0, ac2, bced}.
\subsection{The Properties of the Avian Compass}
There are also two important properties of the avian compass that can be explained by radical pair mechanism very well.
\\
\\
First of all, the avian compass is an inclination compass rather than a polarity compass \cite{ww00, ww1, ww0, nmmh}. The geomagnetic field lines leave the earth at the magnetic pole near the geographic south pole, curve around the earth and re-enter to the earth at the magnetic pole near the geographic north pole. The inclination, defined as the angle between the field lines and the horizontal plane, is $90^{\circ}$ at the magnetic poles and $0^{\circ}$ at the magnetic equator; at the latter, the field lines run parallel to the earth's surface. In the northern hemisphere, field lines have a downward component with a positive inclination; in the southern hemisphere they have an upward component. Therefore, the geomagnetic field provides a reliable, omnipresent source of navigational information for birds capable of sensing the field. The functional mode of the avian magnetic compass is different from that of our standard compasses. The  avian magnetic compass was found to be an inclination compass based on the inclination of the field lines instead of their polarity \cite{ww00}. Birds can only perceive the axial course of the field lines and they must interpret the inclination of the field lines with respect to up and down to derive unambiguous directional information. The avian magnetic compass does not distinguish between magnetic ``north'' and ``south" as indicated by polarity. 
\\
\\
A second important property of the avian compass is its light dependence. This property has been supported by a series of behavioral experiments \cite{ww2,ww3,ww4,ww5,ww6,ww7, ww8, mcd}. The avian compass orientation is dependent on the wavelength of the ambient light. While European robins and Australian silvereyes demonstrated good orientation in blue light and green light, they were disoriented in red light \cite{cc0, ww2, ww3}. Also, European robins cannot orient in yellow-orange light, although there is a significant overlap between the green light, where orientation was excellent, and the yellow light without orientation \cite{ww5}. These experimental findings put constraints on any mechanism designed to explain magnetoreception in birds and provide very strong support for the radical pair mechanism \cite{mcb, edm}.
\subsection{Basic Scheme of the Radical Pair Mechanism}
The actual reaction scheme of the magnetoreception can be very complex in birds. It involves molecules in the eyes absorbing the light, the chemical reaction of the radical pair in these molecules, and the avian neurons to react to the signal. However, we can simplify the scheme into three stages.
\\
\\
In the first stage, the photons with enough energy activate a certain type of molecules located in the birds' eye, inducing an electron transfer reaction and generating the radical pairs in their excited singlet states. After the pair is generated, under the influence of the external magnetic field (the geomagnetic field) and the internal magnetic field (the hyperfine coupling effect), the state of the pair can remain a singlet state or become a triplet state. A different inclination angles, associated with the external magnetic field, can induce different ratios of the singlet and triplet states. In the last stage, the molecules in different states will generate different chemical products which can induce a detectable signal that the birds can use to recognize the direction they need to go \cite{rsd}. 
\\
\\
The predictions of the radical-pair mechanism are consistent with behavioral findings: the reaction yield is independent of the polarity of the magnetic field \cite{ww1,yd0}, reception of magnetic information takes place in the eye \cite{lmc, mdi}, reception is strongly affected by the ambient light conditions \cite{ww2, ww3, ww4, ww5, ww7}, and reception is consistent with the postulated role of ocular photoreceptors in creating magnetosensitive radical pairs \cite{mcb}.
\\
\\
Although there are not any observations of specific biological radical pairs satisfying all constraints, the current discussion of cryptochrome 1a (Cry1a) makes it a promising candidate for magnetosensitive radical pair\cite{rsd, cry1,cry2,cry3, emf, acv, cmb}. Also, the electron transfer path has been studied \cite{et1,et2}. However, the neural path is still unknown, although a vision-based hypothesis was put forward by Ritz $et$ $al$. \cite{cc0}.
\subsection{Reaction Kinetics in Cryptochrome}
Using a carotenoid-porphyrin-fullerene model system \cite{sa2}, Hore and his colleagues demonstrated that, under a static magnetic field, as weak as the geomagnetic field, a chemical reaction can act as a magnetic compass, by producing the detectable changes in the chemical product yield. Their experiment has provided a proof-of-principle of a chemical compass. Along with other experiments \cite{ww1,ww2,ww3,ww4,ww5, ww6}, this provides a significant support to the radical-pair hypothesis of the avian magnetic compass. 
\\
\\
However, the candidate molecules with certain biophysical characteristics must exist in the eyes of the migratory birds, so that the radical pair mechanism can play a role in the magnetoreception. Cryptochromes, a class of photoreceptor proteins located in the retina and absorbing blue-green part of the spectrum which is the functional range of the magnetic compass \cite{rcm}, were proposed as the host molecules for the crucial radical pair cofactors that putatively act as primary magnetoreceptors. Also, the experiments have demonstrated that the vertebrate cryptochromes, under the UV/visible transient absorption and the electron paramagnetic resonance spectroscopy, form long-lived radical pairs which involve a flavin radical and a radical derived from a redox-active amino acid \cite{cry4, cry5}. This demonstrates that the cryptochrome harbors are the type of radical pair needed for the action of the magnetic compass. Scientists also have verified that cryptochromes exist in the eyes of the migratory birds \cite{rcm,cry1,cry4, cry6,cry7}. Furthermore, a distinct part of the forebrain, which primarily processes input from the eyes, is highly active at night in the night-migratory birds \cite{fb1,fb2,fb3,fb4, fb5}. All of these findings are consistent with the hypothesis that the cryptochromes can serve as the primary magnetoreceptors. 
\\
\\
The process of cryptochrome photoactivation has been discussed, and several reaction schemes have been proposed \cite{mia, et1, cry6, rs1,rs2,rs3,rs4, rs5, rs6, rs7, rs8, timescale, arpc}. Cryptochrome contains a blue-light-absorbing chromophore, flavin adenine dinucleotide (FAD). This FAD cofactor is reduced via a series of light-induced electron transfers, from a chain of three tryptophans (Trp) that bridge the space between FAD and the protein surface \cite{timescale}.
\\
\\
According to Hore and his colleagues \cite{et2}, due to the light-induced electron transfer, a radical pair [FAD$^\bullet$$^{-}$TrpH$^\bullet$$^{+}$] (RP1) is formed in the cryptochrome (Cry-1 from the plant \emph{Arabidopsis thaliana}, \emph{At}Cry), followed by a second radical pair [FADH$^\bullet$Trp$^\bullet$] (RP2). The formation of RP2 is more complex. The tryptophanyl radical may deprotonate either fully or partially, while the protonation of FAD$^\bullet$$^{-}$ produces a the neutral FADH$^\bullet$. In this scheme, the RP1 interconverts coherently between singlet and triplet states, under the influence of the magnetic interactions internal to the radicals (electron-nuclear hyperfine couplings), and under the Zeeman interactions with the external magnetic field. Only the singlet state of RP1 can revert to the ground state [FAD+TrpH] by the electron-hole recombination, and the corresponding reaction of the triplet state being spin-forbidden. Simultaneously, one of the constituents of the RP1 changes its protonation state to give RP2, a process that is not spin-related, and in which the singlet and triplet are produced at equal rates.
\subsection{Applications Inspired by the Radical Pair Mechanism}
Despite all of the theoretical arguments, the ultimate goal of studying the mechanisms of bird navigation is to learn from nature and to design highly effective devices that can mimic biological systems in order to detect weak magnetic fields, and to use the geomagnetic field to navigate. Previous literature has shown that the anisotropic hyperfine coupling plays a crucial role in the magnetic field sensitivity of the avian compass and has provided routes towards the design of biologically inspired magnetic compass sensors \cite{mcm, ssrk, sa0, sa4, sa5, ccm, enq, ehc, mnf,osr}. Scientists have exploited many practical methods to realize a device based on the avian chemical compass. Recently, several intriguing models were reported, and significant experimental/simulative results were obtained. 
\\
\\
One of the models is a synthetic donor-bridge-acceptor compass, which is a triad composed of linked carotenoid (C), porphyrin (P) and fullerene (F) groups \cite{sa1,sa2, ssrk, pccn}. Such a triad molecule is the first known molecule that has been experimentally demonstrated to be sensitive to the geomagnetic field, although it works at low temperature (193K). In such a molecule, both radicals are immobilized. The anisotropic magnetic interactions are preserved because the hyperfine interactions in the fullerence radical are very small, and the anisotropic interaction in the carotenoid radical dominates the anisotropy of magnetic field effect. The other conditions are that the radical pair is generated in the singlet state with different decay rates for singlet and triplet states ($k_T<k_S$), and the rate of the spin-lattice relaxation is comparable to the combination rate. Based on all such conditions, the experiments showed a promising result under a magnetic field of the order of $\sim$mT and produced the surprising result that the lifetime of the radical pair is significantly affected by magnetic fields as weak as 50 $\mu$T, which is about the magnitude of the geomagnetic field\cite{sa2}. Yet, the fact is that currently there is not a biomimetic or synthetic chemical compass that functions at room temperature.
\\
\\
Another model is one utilizing magnetic nanostructures to design a chemical compass inspired by theoretical studies \cite{sa3}. According to the theoretical and experimental studies, the internal anisotropic magnetic field is the key factor for product yields of the photochemical reactions to be sensitive to the angle with respect to the external magnetic field \cite{cc0,ac2,sa2}. In some molecules, the hyperfine coupling provides the internal anisotropic magnetic field. However, when designing a real device, the hyperfine coupling can be replaced by a local strong gradient magnetic field, which can be created in the vicinity of a hard ferromagnetic nanostructure by applying a spatially uniform bias field that cancels the field of the nanostructure in a small region \cite{nano,sa3}. In simulations, such a device exhibits significant directional sensitivity \cite{sa3}.
\\
\\
Other studies have achieved to map nanomagnetic field using a radical pair reaction \cite{mnf} and found a method to improve the magnetic sensitivity of chemical magnetometers \cite{osr}.
\subsection{Quantum Coherence and Entanglement in Avian Compass}
Biological mechanisms, such as magnetoreception in birds, are powered by the chemical machinery, which consists of complex molecules structured at the nanoscale and sub-nanoscale. The dynamics of the chemical machinery at such small scales is regulated by the laws of quantum mechanics.  As one of the main features of complex quantum systems, the quantum coherence between different sub-systems, can ultimately lead to many collective phenomena, such as the quantum entanglement. 
	\subsubsection{Effects of Noise}
		\paragraph{Thermal Noise}
	Quantum effects, such as coherence and entanglement, are easily destroyed by interactions with the environment. There is a rule of thumb, the $k_BT$ argument, stating that when the energy of the interaction is smaller than room temperature, then quantum coherent effects cannot persist. Intuitively, because the temperature of the biological systems is around 300$K$, which is hot, thermal fluctuations will destroy the quantum coherent effects. However, the $k_BT$ argument can break down when dealing with living systems like birds. The thermal argument is true only for the equilibrium state, which the system approaches for relatively long times. However, if the equilibrium is approached rapidly, coherent quantum effects can survive (at equilibrium) even at room temperatures \cite{ps1,ps2,ps3,ps4,ps5,ps6, ps7, netq, aslq}. Another possibility is that the quantum coherent effects, including the quantum entanglement, can survive as the system approaches the  equilibrium with the surrounding environment. In another word, the quantum effects are used before the system has time to equilibrate with the environment. The latter situation is believed to be what happens in the radical pair mechanism of the avian compass \cite{ER}.
\\
\\
Since the geomagnetic field is very weak ($\approx 0.5$ $Gauss$), the effects of such a weak magnetic field, in biological systems, can easily be masked by the thermal fluctuations. To take advantage of the weak geomagnetic field, one effective method is to use the short-lived, specialized photochemical reactions. Then the thermal fluctuations do not have enough time to effectively mask the effects of the weak magnetic fields. The radical pair mechanism is believed to rely on this process \cite{cry2}.
\\
\\
If the useful signal carries more energy than the thermal noise, this signal can be distinguished from the noise. But since the energy differences, due to the geomagnetic fields, between the singlet and the triplet states are generally small, and are even smaller in living organisms, the chemical reaction of radical pairs is a subject to a significant thermal noise \cite{tn1}. However, due to the short lifetime  of the radical pair reactions, and due to the fast conversion rate between singlet and triplet states, the photochemical reactions can produce a bird-detectable signal through the yields of singlet and triplet states, in spite of the thermal noise. Weaver $et$ $al$. also discussed the detection limit in the radical pair mechanism, and claimed that a chemical compass in birds is feasible with the chemical reactions with certain rate constants \cite{tn2}.
\\
\\
		\paragraph{Dephasing Noise}
Other than thermal noise, dephasing noise can only affect the coherences between singlets and triplets. Once the radicals come close to each other, dipolar and exchange interaction can no longer be neglected. For instance, if the exchange interaction is considered, the electron singlet and triplet states will be separated in energy, which can destroy the conversion between the singlet and the triplet \cite{dcc}. However, since the molecule motion can be quickly distributed over the environment owing to Brownian molecule motion, the microscopic causes of the path taken are quickly lost. In addition, the involved energy scales are much smaller when compared with the energy released owing to recombination\cite{dcc}. Therefore, the dephasing noise can hardly harm the coherence since the molecular motions that cause spin-decoherence by modulating hyperfine interactions have either low amplitude or high frequency or both.
\\
\\
There are many papers that discuss the effects of dephasing noise on the radical pair mechanism via Lindblad operators \cite{dcc, sqc, sa4}. Intuitively, one might expect that dephasing noise is unfavorable for the function of the chemical compass. However, these studies show that the effect of noise depends on the model. In Gauger's model, the compass mechanism is almost immune to pure phase noise \cite{sqc}. Cai argued that correlated dephasing noise could even enhance the chemical compass in their model \cite{sa4}. Both models give us positive, or at least non-negative, perspectives concerning dephasing noise. 
	\subsubsection{Quantum Coherence}
	As we know quantum coherence describes a particle's ability to exist simultaneously in several distinct states, such as position, energy, or spin. Quantum coherence plays a crucial role in the energy transport in photosynthetic complexes \cite{ps5, ps8, ps10, spec, rqce, mshe, vsle, cdbs, cooc}. There is indirect evidence indicating that the quantum coherence exists in bird navigation \cite{cc0, ac5}. The proposed radical pair mechanism for avian compass postulates that absorption of blue or green light creates a radical pair of electrons. The radical pair undergoes coherent quantum oscillations between entangled singlet and triplet states, at a rate depending on the external magnetic fields. Finally, the pairs in the singlet and the triplet states will lead to different chemical products, providing orientation information for birds. So, the quantum coherence plays an important role in this mechanism.
	\subsubsection{Quantum Entanglement}
	Quantum entanglement is the quantum mechanical property that allows two or more particles to be correlated differently and stronger than the same classically correlated particles.  This also means that while the global state of the system is perfectly known, the local state is fully mixed. The entanglement creates the non-local correlations, and the non-thermal excitations \cite{ER, quen, ref1}.
	\\
	\\
	Quantum entanglement plays a key role in the radical pair mechanism of the avian compass. A single electron photo-excitation and a subsequent electron translocation leads to an entangled state, which provides the necessary spin correlations. Along with the effects of the magnetic fields, this causes the electrons to oscillate between singlet and triplet states. 
\\
\\
Many scientists have conducted researches on the role of entanglement in chemical compass \cite{et4, sqc, esma}. Gauger $et$ $al$ shown that even though entanglement falls off at a faster rate than the decay of population from the excited state under dephasing noise, the superposition and entanglement are sustained for at least tens of microseconds, exceeding the durations achieved in the best comparable man-made molecular system \cite{sqc}. However, other researchers focus on the effects of the entanglement of the initial states of the radical pairs\cite{et4,esma}. These studies have shown that the sensitivity of the chemical compass depends on the initial state of the radical pair. Briegel and his colleagues have randomly chosen 5000 different initial states from the set of separable states and singlet initial state and calculated the maximal achievable magnetic-field sensitivity for every value of $B$. The results show that entanglement is indeed helpful, and it is specifically entanglement rather than mere quantum coherence \cite{et4}.
	\\
	\\
	``Negativity", originated from the observation due to Peres \cite{peres}, is a computable measure of the entanglement. It is based on the trace norm of the partial transpose, $\rho^{T_A}$, of the bipartite mixed state density matrix, $\rho$ \cite{ccsd,GR}. Essentially, negativity measures the degree to which $\rho^{T_A}$ fails to be positive, and therefore it can be regarded as a quantitative version of Peres criterion of separability \cite{peres,ccsd, GR}. From the definition of negativity, we can construct a quantity to measure the entanglement: $N(\rho)=\frac{\| \rho^{T_A} \|_1-1}{2}$, where $\| \rho^{T_A} \|_1$ is the trace norm of the partial transpose of the system's density matrix  \cite{GR,kais1}. This measure of entanglement, $N(\rho)$, corresponds to the absolute value of the sum of negative eigenvalues of $\rho^{T_A}$\cite{bb,lv}, and it vanishes for unentangled states. Moreover, in the case of an entangled state of two qubits, negativity is defined as two times the absolute value of the negative eigenvalue of the partial transpose of a state \cite{fv}.
\\
\\
	Denote $\rho$ as a generic state of a bipartite system with the finite-dimensional Hilbert space, $\mathcal {H}_A \otimes \mathcal{H}_B$, shared by two parties, A and B. $\rho^{T_A}$, the partial transpose of the density matrix, $\rho$, with respect to subsystem A, will be defined as: 
	\begin{equation}
	\langle i_A, j_B\mid\rho^{T_A}\mid m_A, n_B\rangle \equiv \langle m_A, j_B\mid\rho\mid i_A, n_B\rangle ,
	\end{equation}
	 where $\mid i_A, j_B\rangle \equiv \mid i\rangle_A \otimes \mid j\rangle_B \in \mathcal {H}_A \otimes \mathcal{H}_B$ is a fixed but arbitrary orthonormal product basis \cite{GR}. The trace norm of $\rho^{T_A}$ is: 
	 \begin{equation}
	 \parallel \rho^{T_A} \parallel_1 = \text{tr}\sqrt{{\rho^{T_A}}^{\dagger}\rho^{T_A}},
	 \end{equation}
	 which is equal to the sum of the absolute values of the eigenvalues of $\rho^{T_A}$, since $\rho^{T_A}$ is hermitian \cite{mb}. Since the eigenvalues of the density matrix, $\rho$, are positive, the trace norm of $\rho$ is: $\parallel\rho\parallel_1=\text{tr}\rho=1$. Thus, the partial transpose, $\rho^{T_A}$, also satisfies the condition: tr$[\rho]=1$. But since it may have negative eigenvalues, $e_i < 0$, its trace norm is: $\parallel \rho^{T_A} \parallel_1=1+2\mid\sum_ie_i\mid\equiv 1+ 2 N(\rho)$ \cite{GR}. Therefore, the negativity ($N(\rho)=|\sum_ie_i|$) can also be defined as the sum of the negative eigenvalues, $e_i$, of the density matrix partial transpose, $\rho^{T_A}$, measuring by how much $\rho^{T_A}$ fails to be positive definite \cite{GR, wg}. So we have an alternative way to calculate the negativity. It can be written as:
	 \begin{equation}
	 N(\rho)=\mid\sum_ie_i\mid=\sum_j\frac{\mid\lambda_j\mid-\lambda_j}{2},
	 \end{equation}
	 where $e_i$ are the negative eigenvalues of $\rho^{T_A}$, and $\lambda_j$ are the eigenvalues of $\rho^{T_A}$. In the following, we will show some simple examples to illustrate negativity.
	\\
	\\
	Assuming the state $|\psi\rangle$ can be separated into two sub-states, $|\phi_A\rangle$ and $|\phi_B\rangle$, indicating that this state is not an entangled state in which $|\phi_A\rangle = \frac{1}{\sqrt{2}}(|\uparrow\rangle + |\downarrow\rangle)$ and $|\phi_B\rangle = \frac{1}{\sqrt{2}}(|\uparrow\rangle + |\downarrow\rangle)$, the expression of $|\psi\rangle$ will be:
	\begin{equation}
	|\psi\rangle = |\phi_A\rangle\otimes |\phi_B\rangle =\frac{1}{2}(|\uparrow\rangle_A + |\downarrow\rangle_A) \otimes (|\uparrow\rangle_B + |\downarrow\rangle_B).
	\end{equation}
	As a result, the density matrix of $|\psi\rangle$, $\rho_\psi$, will be:
	\begin{equation}
        \begin{split}
        \rho_\psi &= |\psi\rangle\langle\psi|\\
        &=\frac{1}{4}\left(\begin{array}{cccc}
	$1$ & $1$ & $1$ & $1$\\
	$1$ & $1$ & $1$ & $1$\\
	$1$ & $1$ & $1$ & $1$\\
	$1$ & $1$ & $1$ & $1$\\
	\end{array}\right) .
        \end{split}
	\end{equation}
	Therefore, the partial transpose of the density matrix with respect to subsystem $B$ is, 
	\begin{equation}
        \begin{split}
        \rho_\psi^{T_B} =\frac{1}{4}\left(\begin{array}{cccc}
	$1$ & $1$ & $1$ & $1$\\
	$1$ & $1$ & $1$ & $1$\\
	$1$ & $1$ & $1$ & $1$\\
	$1$ & $1$ & $1$ & $1$\\
	\end{array}\right) .
        \end{split}
	\end{equation}
	Through the calculation, we can find that the four eigenvalues of $\rho_\psi^{T_A}$ are $\{1,0,0,0\}$. According to the formula of negativity above, $N(\rho) = \sum_i^4\frac{\mid\lambda_i\mid-\lambda_i}{2}$, we find that the negativity is $0$. This example shows us that the negativity of a unentangled state is zero.
	\\
	\\
	Next, we will consider an example of an entangled state $|\psi\rangle=\frac{1}{\sqrt{2}}(\mid\uparrow\downarrow\rangle-\mid\downarrow\uparrow\rangle)=\frac{1}{\sqrt{2}}(0\ 1\ 1\ 0)^{T}$. In this case, the density matrix of the state $\mid\psi\rangle$, $\rho_\psi$, is
	\begin{equation}
        \begin{split}
        \rho_\psi =\frac{1}{2}\left(\begin{array}{cccc}
	$0$ & $0$ & $0$ & $0$\\
	$0$ & $1$ & $-1$ & $0$\\
	$0$ & $-1$ & $1$ & $0$\\
	$0$ & $0$ & $0$ & $0$\\
	\end{array}\right) .
        \end{split}
	\end{equation}
	Therefore, the partial transpose of this density matrix with respect to subsystem $B$ is 
	\begin{equation}
        \begin{split}
        \rho_\psi^{T_B} =\frac{1}{2}\left(\begin{array}{cccc}
	$0$ & $0$ & $0$ & $-1$\\
	$0$ & $1$ & $0$ & $0$\\
	$0$ & $0$ & $1$ & $0$\\
	$-1$ & $0$ & $0$ & $0$\\
	\end{array}\right) .
        \end{split}
	\end{equation}
	We can find that the eigenvalues of $\rho_\psi^{T_B}$ are $\{-0.5,0.5,0.5,0.5\}$. Therefore, the negativity of the state $|\psi\rangle$ is $0.5$. After taking twice the negativity, we find that the entanglement of the state $|\psi\rangle$ is 1, which is the maximum entanglement.
\\
\\
The above examples show that negativity can serve as a metric of the entanglement. Therefore, in the following calculations we will use the negativity as the measurement of entanglement. However, it is worth to mention that there is another measurement of the entanglement named concurrence \cite{amec}. For a two-qubit system, the two measurements, negativity and concurrence, are equivalent.

\section{Calculations and Results}
\begin{figure}
\begin{center}
\includegraphics[width=0.75\textwidth]{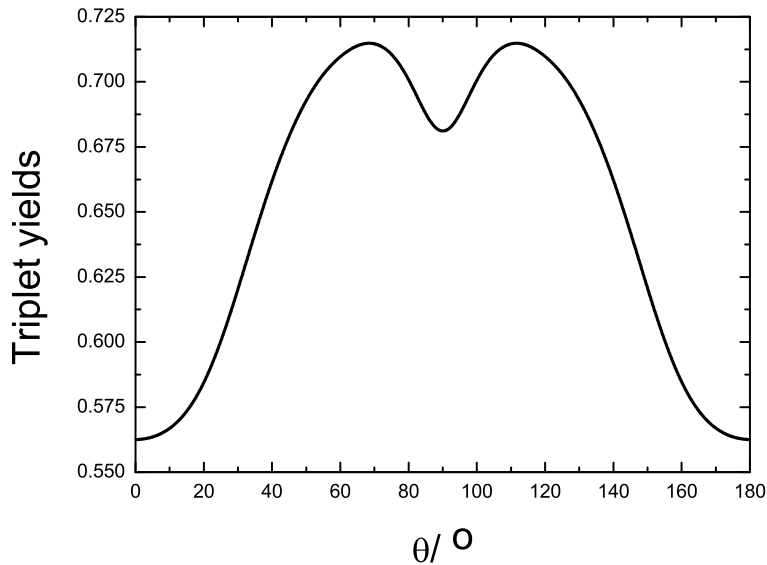}
\end{center}
\caption{\small {Angular dependence of the triplet yields. The triplet yields are symmetric around $90^\circ$. }}
\label{yields}
\end{figure}
In this section, we will introduce some of our work on the topic of the radical pair mechanism \cite{work, et3}. We investigate two aspects of the radical pair mechanism: the yields of defined signal states and the entanglement of the states.
\subsection{Basic One-Stage Scheme}
Following Ref. \cite{cc0}, we include only the Zeeman interaction and the hyperfine interaction in the Hamiltonian of the system:
\begin{equation}
H=g \mu_B \sum_{i=1}^2 \vec S_i \cdot \left( \vec B + \widehat A_i \cdot \vec I_i \right) \label{1sh}.
\end{equation}
In Eq. (\ref{1sh}), the first term is the Zeeman interaction and the second term is the hyperfine interaction. (We assume that each electron is coupled to a single nucleus.) $\vec I_i$ is the nuclear spin operator; $\vec S_i$ is the electron spin operator, i.e., $\vec S={\vec\sigma}/{2}$ with $\vec\sigma$ being the Pauli matrices; $g$ is the $g$-factor of the electron, which is chosen to be $g=2$; $\mu_B$ is the Bohr magneton of the electron; and $\widehat A_i$ is the hyperfine coupling tensor, a 3$\times$3 matrix.
\\
\\
As proposed in Ref. \cite{cc0}, we model the radical-pair dynamics with a Liouville equation \cite{hab},
\begin{align}
\dot{\rho}(t)=&-\frac{i}{\hbar}[H,\rho(t)] \nonumber\\
                    &-\frac{k_S}{2}\left\{Q^S,\rho(t)\right\}-\frac{k_T}{2}\left\{Q^T,\rho(t)\right\} \label{1sl}.
\end{align}
In Eq. (\ref{1sl}), $H$ is the Hamiltonian of the system; $Q^S$ is the singlet projection operator, i.e. $Q^S=|S\rangle\langle S|$, and $Q^T=|T_+\rangle\langle T_+|+|T_0\rangle\langle T_0|+\mid T_-\rangle\langle T_-|$ is the triplet projection operator, where $|S\rangle$ is the singlet state and ($|T_+\rangle, |T_0\rangle, |T_-\rangle$) are the triplet states \cite{Lambert}; $\rho(t)$ is the density matrix for the system; $k_S$ and $k_T$ are the decay rates for the singlet state and triplet states, respectively.
\\
\\
Under the basic scheme, we assume that the initial state of the radical pair is a perfect singlet state, $\mid$S$\rangle=\frac{1}{\sqrt{2}}(\mid\uparrow\downarrow\rangle-\mid\downarrow\uparrow\rangle)$. Therefore, the initial condition for the density matrix is: $\rho(0)=\frac{1}{4}\hat{I}_N\otimes {Q}^S$, where the electron spins are in their singlet states, and nuclear spins are in a completely mixed state, which is a 4$\times$4 identity matrix. Assuming that the rate is independent of spin, the decay rates for the singlet and triplet should be the same \cite{cc0}, $k_S=k_T=k=1\mu$s$^{-1}$, i.e., $k$ is the recombination rate for both the singlet and triplet states. The external weak magnetic field, $\vec{B}$, representing the Earth's magnetic field in Eq. (\ref{1sh}), depends on the angles, $\theta$ and $\varphi$, with respect to the reference frame of the immobilized radical pair, i.e., $\vec{B}=B_{0}(\sin\theta\cos\varphi, \sin\theta\sin\varphi, \cos\theta)$, where $B_0=0.5$G is the magnitude of the local geomagnetic field. Without losing the essential physics, $\varphi$ can be assumed to be $0$.
\\
\\
Since the radical pair must be very sensitive to different alignments of the magnetic field, it is necessary to assume that the hyperfine coupling tensors in Eq. (\ref{1sh}) are anisotropic. However, for the sake of simplicity, we employ the hyperfine coupling as anisotropic for one radical, and as isotropic for the other \cite{cc0}, i.e.,
\begin{displaymath}
\widehat{A_1}=
\left(\begin{array}{ccc}
$10G$ & $0$ & $0$\\
$0$      &$10G$ &$0$ \\
$0$& $0$& $0$
\end{array}\right)
\normalsize{,}
\
\widehat{A_2}=
\left(\begin{array}{ccc}
$5G$ & $0$ & $0$\\
$0$      &$5G$ &$0$ \\
$0$& $0$& $5G$
\end{array}\right).
\end{displaymath}
Using the parameters defined above, we calculate one of the properties of the avian compass, depending only on the inclination but not on the polarity. As one can see in Fig. \ref{yields}, the triplet yields are symmetric around $90^\circ$ since the hyperfine coupling tensors are symmetric. Consequently, the radical pair mechanism cannot distinguish between magnetic fields that are oppositely directed but have the same magnitude. 
\begin{figure}
\begin{center}
\includegraphics[width=0.75\textwidth]{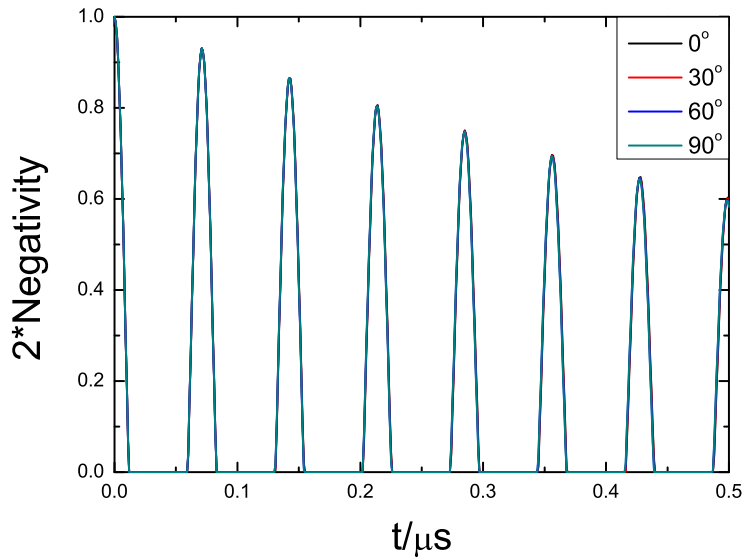}
\end{center}
\caption{\small{Entanglements for different angles. Results are produced by assuming the decay rate to be $1\mu$s$^{-1}$. All curves are almost identical. In the geomagnetic field, entanglement does not change with orientation.}}
\label{angle}
\end{figure}
\\
Also, we investigated whether the entanglement, measured by negativity, is angle-dependent. While, using the suggested hyperfine coupling tensor in Ref. \cite{cc0}, the calculation gives us the surprising result shown in Fig. \ref{angle}. Namely, the dynamics of entanglement does not change with angle, i.e., the entanglement is not sensitive to the angle between the $z$-axis of the radical pair and the Earth's magnetic field under these parameters. The main reason may be that the fast exponential decay rate hides the effect of the hyperfine coupling interaction. Therefore, when enlarging the hyperfine coupling, the difference of entanglement between angles will be observable. Even though the entanglement of the radical pair cannot directly provide directional information, this does not mean that entanglement is not involved in avian navigation. There might be indirect mechanisms for birds to utilize entanglement.
\\
\\
The results of our calculations shown in Fig. \ref{angle}, demonstrate that the dynamics of entanglement is almost the same for all angles when symmetric hyperfine tensors are involved. This raises the following question: ``What will happen if there is an asymmetric hyperfine tensor?" Although hyperfine tensors of organic radicals are usually symmetric, we can examine a few asymmetrical cases to try to find the possible asymmetry effects of the hyperfine coupling. The radicals with asymmetric hyperfine tensors are likely to have fast spin relaxation and not suitable for the purpose of a geomagnetic sensor. Therefore, our study here is only out of theoretical curiosity to find out what will happen to the entanglement when the hyperfine tensors are not symmetric. The asymmetric hyperfine tensors we examine are:
\begin{displaymath}
\widehat{A{^b_1}}=
\left(\begin{array}{ccc}
$10G$ & $0$ & $0$\\
$0$      &$10G$ &$0$ \\
$0$& $0$& $4G$
\end{array}\right)
\normalsize{,}
\
\widehat{A{^b_2}}=
\left(\begin{array}{ccc}
$5G$ & $5G$ & $0$\\
$0$      &$5G$ &$0$ \\
$0$& $0$& $5G$
\end{array}\right),
\end{displaymath}
and
\begin{displaymath}
\widehat{A{^c_1}}=
\left(\begin{array}{ccc}
$0$ & $0$ & $0$\\
$0$      &$0$ &$0$ \\
$0$& $0$& $4G$
\end{array}\right)
\normalsize{,}
\
\widehat{A{^c_2}}=
\left(\begin{array}{ccc}
$0$ & $5G$ & $0$\\
$0$      &$0$ &$0$ \\
$0$& $0$& $0$
\end{array}\right).
\end{displaymath}
\begin{figure}
\includegraphics[width=0.5\textwidth]{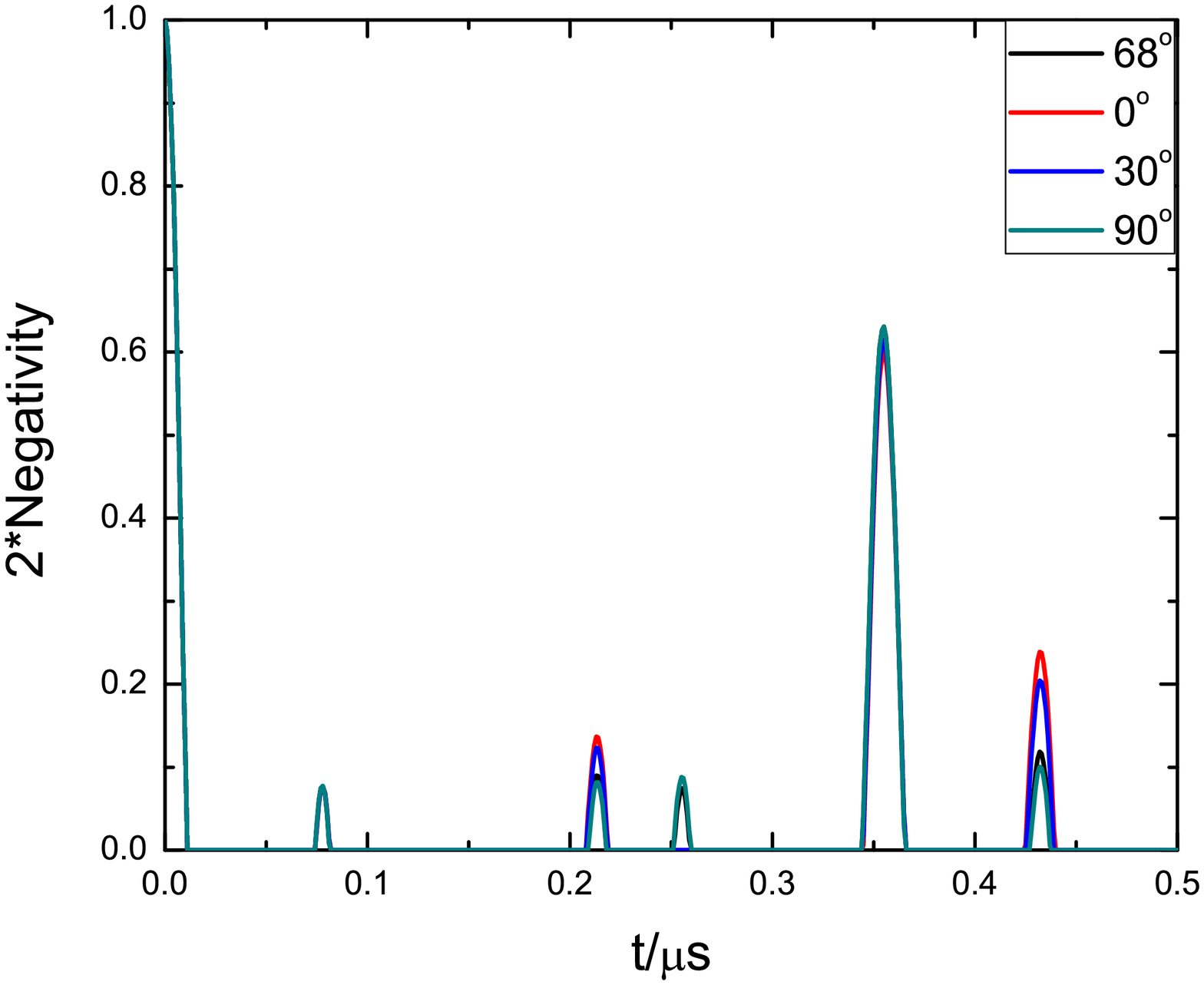}
\includegraphics[width=0.5\textwidth]{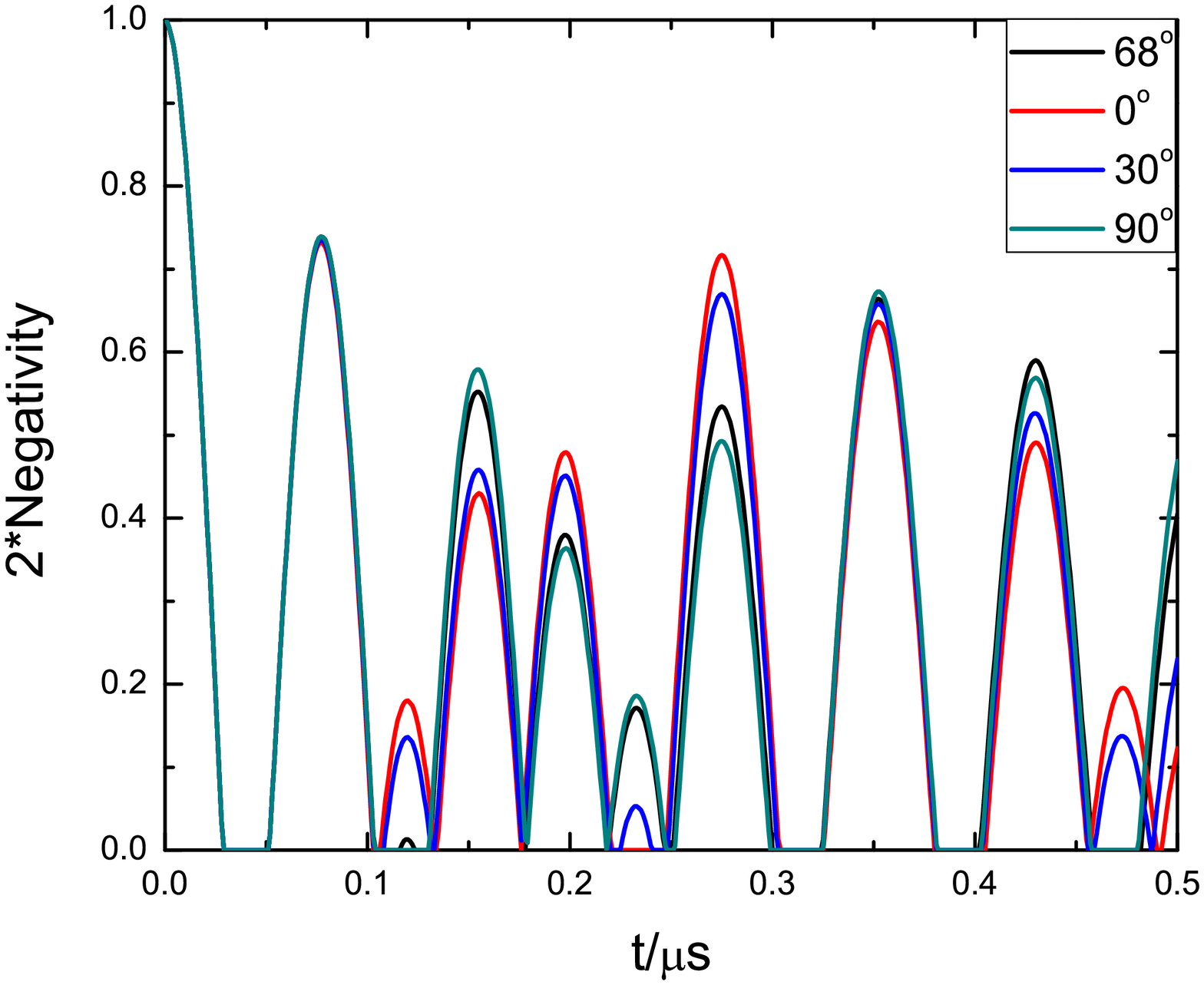}
\caption{\small{Entanglements for four angles under the hyperfine coupling tensors, $\widehat A{^b_i}$ and $\widehat A{^c_i}$.}}
\label{angle_12}
\end{figure}
\\
From Fig. \ref{angle_12}, we can easily see that the hyperfine coupling tensor pair of $\widehat A{^c_i}$ gives an intriguing result. Namely, the dynamics of the entanglement is clearly dependent on the system's orientation.
\subsection{Two-Stage Scheme}
\begin{figure}[!ht]
\begin{center}
\includegraphics[height=2in]{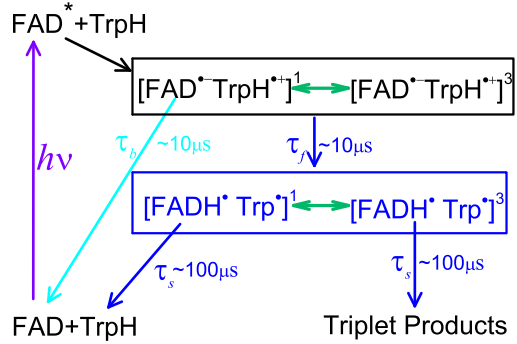}
\end{center}
\caption{\small{The reaction scheme of the radical pair mechanism in cryptochrome. $k_b=\tau_b^{-1}$ and $k_f=\tau_f^{-1}$ are the first-order rate constants for recombination of the initial radical pair and formation of the secondary pair from the initial one, respectively. $k_s=\tau_s^{-1}$ is rate constant for the decay of the secondary pair. The green two-headed arrows indicate the interconversion of the singlet and triplet states of the radical pairs.}}
\label{scheme}
\end{figure}

For further study, we modify our model based on Ref. \cite{et2}. The radical pair reaction scheme has two stages (see Fig. \ref{scheme}). The initial radical pair [FAD$^\bullet$$^{-}$TrpH$^\bullet$$^{+}$] is formed by light-induced electron transfer, followed by the protonation and deprotonation, forming a secondary radical pair [FADH$^\bullet$Trp$^\bullet$]. This two-stage scheme is shown in Fig. \ref{scheme}. Both radical pairs are affected not only by the external magnetic field, but also by their surrounding nuclei. Respectively, the Hamiltonians of the initial and secondary radical pair are,
\begin{equation}
H_1=g \mu_B \sum_{i=1}^2 \vec S_i \cdot \left( \vec B + \widehat A_{1i} \cdot \vec I_i \right), \label{eq1}
\end{equation}
\begin{equation}
H_2=g \mu_B \sum_{i=1}^2 \vec S_i \cdot \left( \vec B + \widehat A_{2i} \cdot \vec I_i \right). \label{eq2}
\end{equation}
\\
In Eq. (\ref{eq1}) and Eq. (\ref{eq2}), $\vec S_i$ is the unpaired electron spin of the radical pairs, and $\vec I_i$ is the nuclear spin of nitrogen in the pairs. We calculate the hyperfine coupling tensors (Table. 1), $\widehat A_{ij}$, using Gaussian09 with UB3LYP/EPR-II. For simplicity, in our subsequent calculations, we only use one of the hyperfine coupling tensors within each molecule, since additional nuclear spins have little effect on the yield curves\cite{et1}. Also, because the electron is located near the nitrogen atoms, and the couplings between the electron and the nitrogen atoms are stronger than the couplings to other near-by hydrogen atoms, we choose, for our subsequent calculations, the hyperfine coupling tensors associated with the nitrogen atoms in each molecule. $\vec B$ is the weak external geomagnetic field. $\vec{B}$ depends on the angles, $\theta$ and $\varphi$, with respect to the reference frame of the immobilized radical pair, i.e., $\vec{B}=B_{0}(\sin\theta\cos\varphi, \sin\theta\sin\varphi, \cos\theta)$. We can choose the x-axis so that the azimuthal angle, $\varphi$, is 0. The constants, $g$ and $\mu_B$, are the $g$-factor and the Bohr magneton of the electron, respectively.
\begin{center} \small
	\begin{tabular}{| l | l | p{2.5cm} | p{3 cm} | p{2.8cm}|}
	\multicolumn{5}{l}{Table. 1: The hyperfine coupling tensors of some atoms.} \\
	\hline
	Moleclule & Atom & Isotropic (G) & Anisotropic (G) & Principal Axis (Angstrom) \\  
	\hline
	TrpH$^\bullet$$^{+}$& $N$ & 12.864 & -7.154 & 0.63  0.70  -0.34 \\
	 & & & -7.051 & 0.73 0.63 0.04\\
	 & & & 14.205 & -0.26 0.22 0.94\\ 
	 & $H$ & -3.054 & -1.478 & 0.50 0.86 -0.12\\
	 & & & -0.977 & -0.14 0.20 0.96\\
	 & & & 2.454 & 0.85 -0.47 0.22\\ 
	\hline
	FAD$^\bullet$$^{-}$ & $N$ & 2.339 & -5.392 & 0.61 0.79 0.00 \\
	 & & & -5.353 & 0.79 0.61 0.00\\
	 & & & 10.745 & 0.00 0.00 1.00\\ 
	 \hline
	Trp$^\bullet$ & $N$ & 8.393 & -9.708 & -0.23  0.97  0.10\\
	 & & & -9.539 & 0.96 0.24 -0.14\\
	 & & & 19.247 & 0.15 -0.07 0.99\\ 
	 & $H$ & -5.888 & -2.458 & 0.62 0.78 -0.09\\
	 & & & -0.792 & 0.21 -0.06 0.98\\
	 & & & 1.320 & 0.75 -0.63 -0.20\\  
	\hline
	FADH$^\bullet$& $N$ & 2.015 & -4.815 & 0.79 0.61 0.00 \\
	 & & & -4.702 & 0.61 0.79 0.00\\
	 & & & 9.517 & 0.00 0.00 1.00\\ 
	 & $H$ & -3.054 & -1.478 & 0.50 0.86 -0.12\\
	 & & & -0.977 & -0.14 0.20 0.96\\
	 & & & 2.454 & 0.85 -0.47 0.22\\  
	\hline
	\end{tabular}
\end{center}
The time evolution of the corresponding spin system is described through a modified stochastic Liouville equation \cite{sle, cc0, ik1, ik2, ik3, ik5}. For this purpose, we denote the density matrix corresponding to the states of the radical pair [FAD$^\bullet$$^{-}$TrpH$^\bullet$$^{+}$] as $\rho_1$ and the density matrix corresponding to the states of the radical pair [FADH$^\bullet$Trp$^\bullet$] as $\rho_2$. Each density matrix follows a stochastic Liouville equation that describes the spin motion and also takes into account the transition into and out of a particular state from or into other states, as illustrated in Fig.\ref{scheme}. Therefore, the dynamics of the radical pairs in the two-stage scheme is governed by the following coupled Liouville equations:
\begin{equation}
\begin{split}
\label{eq3}
\frac{\partial{\rho_1}(t)}{\partial t}=&-\frac{i}{\hbar}[H_1,\rho_1(t)] \\
&-\frac{k_f}{2}\left\{Q^S,\rho_1(t)\right\} -\frac{k_f}{2}\left\{Q^T,\rho_1(t)\right\}\\
&-\frac{k_b}{2}\left\{Q^S,\rho_1(t)\right\},
\end{split}
\end{equation}
\begin{equation}
\begin{split}
\label{eq4}
\frac{\partial{\rho_2}(t)}{\partial t}=&-\frac{i}{\hbar}[H_2,\rho_2(t)] \\
                    &+\frac{k_f}{2}\left\{Q^S,\rho_1(t)\right\}+\frac{k_f}{2}\left\{Q^T,\rho_1(t)\right\}\\
                    &-\frac{k_s}{2}\left\{Q^S,\rho_2(t)\right\}-\frac{k_s}{2}\left\{Q^T,\rho_2(t)\right\},
\end{split}
\end{equation}
where $H_1$ and $H_2$ are the Hamiltonians of the two radical pairs given in Eqs. (\ref{eq1}) and (\ref{eq2}); $Q^S$, as defined before, is the singlet projection operator, $Q^S=|S\rangle\langle S|$, and $Q^T=|T_+\rangle\langle T_+|+|T_0\rangle\langle T_0|+|T_-\rangle\langle T_-|$ is the triplet projection operator, where $|S\rangle$ is the singlet state, and ($|T_+\rangle, |T_0\rangle, |T_-\rangle$) are the triplet states; and all of the decay rates are indicated in Fig. \ref{scheme}. In addition, the initial state of the pair [FAD$^\bullet$$^{-}$TrpH$^\bullet$$^{+}$] is assumed to be in the singlet state, $\mid$S$\rangle=\frac{1}{\sqrt{2}}(\mid\uparrow\downarrow\rangle-\mid\downarrow\uparrow\rangle)$, while the pair [FADH$^\bullet$Trp$^\bullet$] is not produced initially. In other words, $\rho_1(0)=\frac{1}{9}\hat{I}_N\otimes {Q}^S$, where the electron spins are in the singlet states, and nuclear spins are in thermal equilibrium, a completely mixed state, which is a 9$\times$9 identity matrix, and $\rho_2(0)=0$.
\\
\\
We consider the product formed by the radical pair [FADH$^\bullet$Trp$^\bullet$] in the triplet state as the signal product, whose yield is defined as: $\Phi_T=k_s\int_0^\infty Tr[Q^T\rho(t)] dt$ \cite{yd0, yd, ik1}, where $Q^T=|T\rangle\langle T|$, and $\mid$T$\rangle=|T_+\rangle + | T_0\rangle+ | T_-\rangle$.
\begin{figure}[htpb]
\begin{center}
\includegraphics[height=0.6\textwidth]{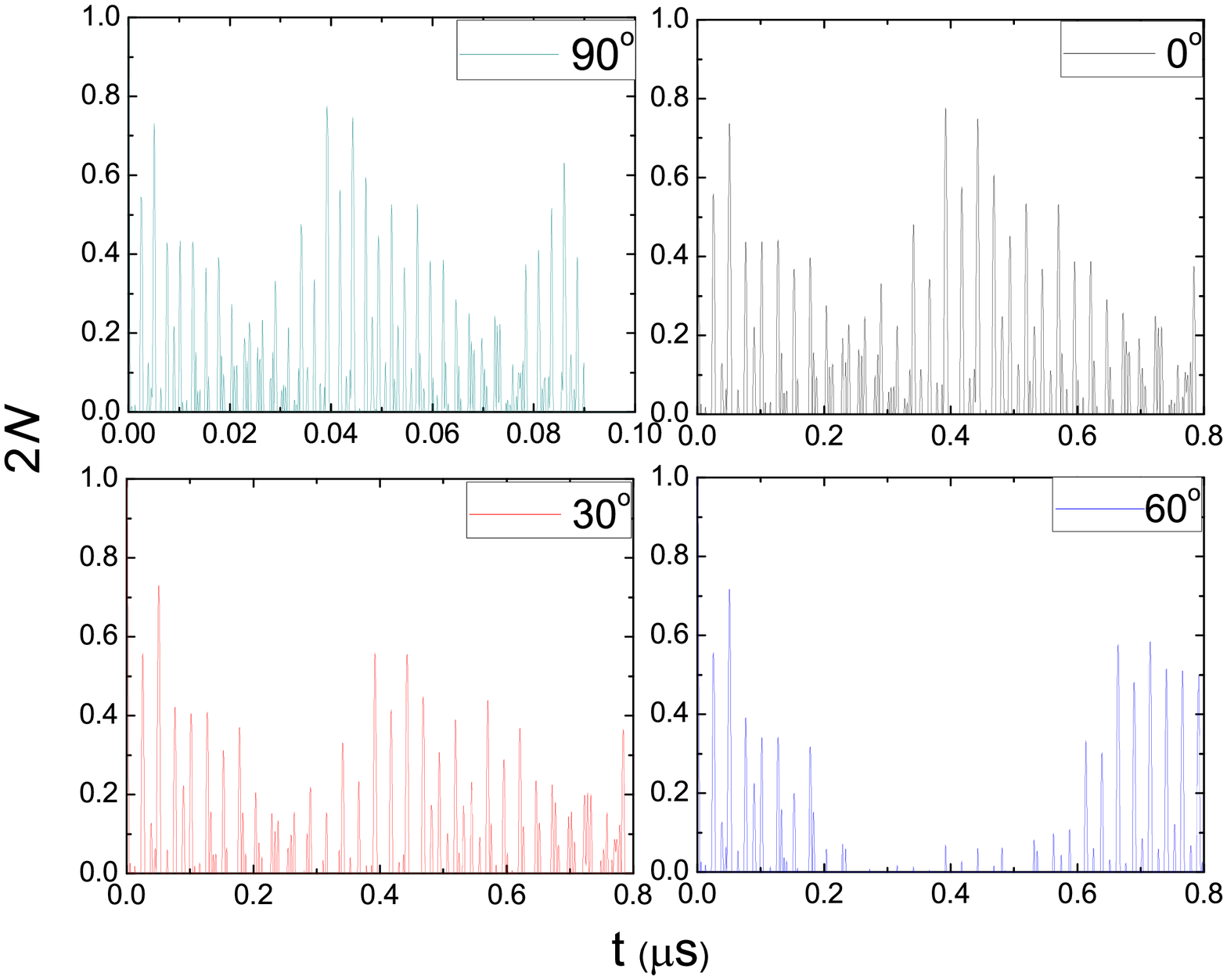}
\end{center}
\caption{\small{Entanglement of the initial radical pair [FAD$^\bullet$$^{-}$TrpH$^\bullet$$^{+}$] as a function of the external fields for four polar angles, $\theta = 0^{\circ}(\text{black}), 30^{\circ}(\text{red}), 60^{\circ}(\text{blue}), 90^{\circ}(\text{green})$. Since the entanglement of the initial pair [FAD$^\bullet$$^{-}$TrpH$^\bullet$$^{+}$] is compressed within 0.1 $\mu$s, the timescale of the graph is from 0 to 0.1 $\mu$s. The other graphs range from 0 to 0.8 $\mu$s. And the entanglements of the initial pair at $0^{\circ}, 30^{\circ}, 60^{\circ}$ differ after 0.3 $\mu$s.}}
\label{entg2}
\end{figure}
\\
The entanglement is believed to play an important role in many systems\cite{kais1,kais2,kais3,kais4}, including the chemical compass in birds. As mentioned before, we use negativity as the metric of entanglement. However, for the two-stage scheme, the secondary radical pair [FADH$^\bullet$Trp$^\bullet$] barely has any entanglement between the two unpaired electrons, since the chemical reaction (protonation and deprotonation) has destroyed the entanglement between them in the preceding radical pair [FAD$^\bullet$$^{-}$TrpH$^\bullet$$^{+}$]. The unpaired electrons in the initial radical pair show a robust entanglement. Fig. \ref{entg2} shows the entanglement of the initial radical pair [FAD$^\bullet$$^{-}$TrpH$^\bullet$$^{+}$] for four polar angles, $\theta$. Also, the dynamics of the entanglement is clearly dependent on the angles, which is very different from the results in the one-stage case \cite{et3}. However, the entanglements at the angles, 0$^{\circ}$, 30$^{\circ}$ and 60$^{\circ}$, are nearly the same for the first 0.1$\mu$s, while the entanglement at 90$^{\circ}$ is very different from others. At 90$^{\circ}$, the entanglement lasts for 0.1$\mu$s, which is long enough for electrons to transfer between different molecules \cite{timescale}.
\begin{figure}[htbp]
	\centering
	\includegraphics[width=0.45\textwidth]{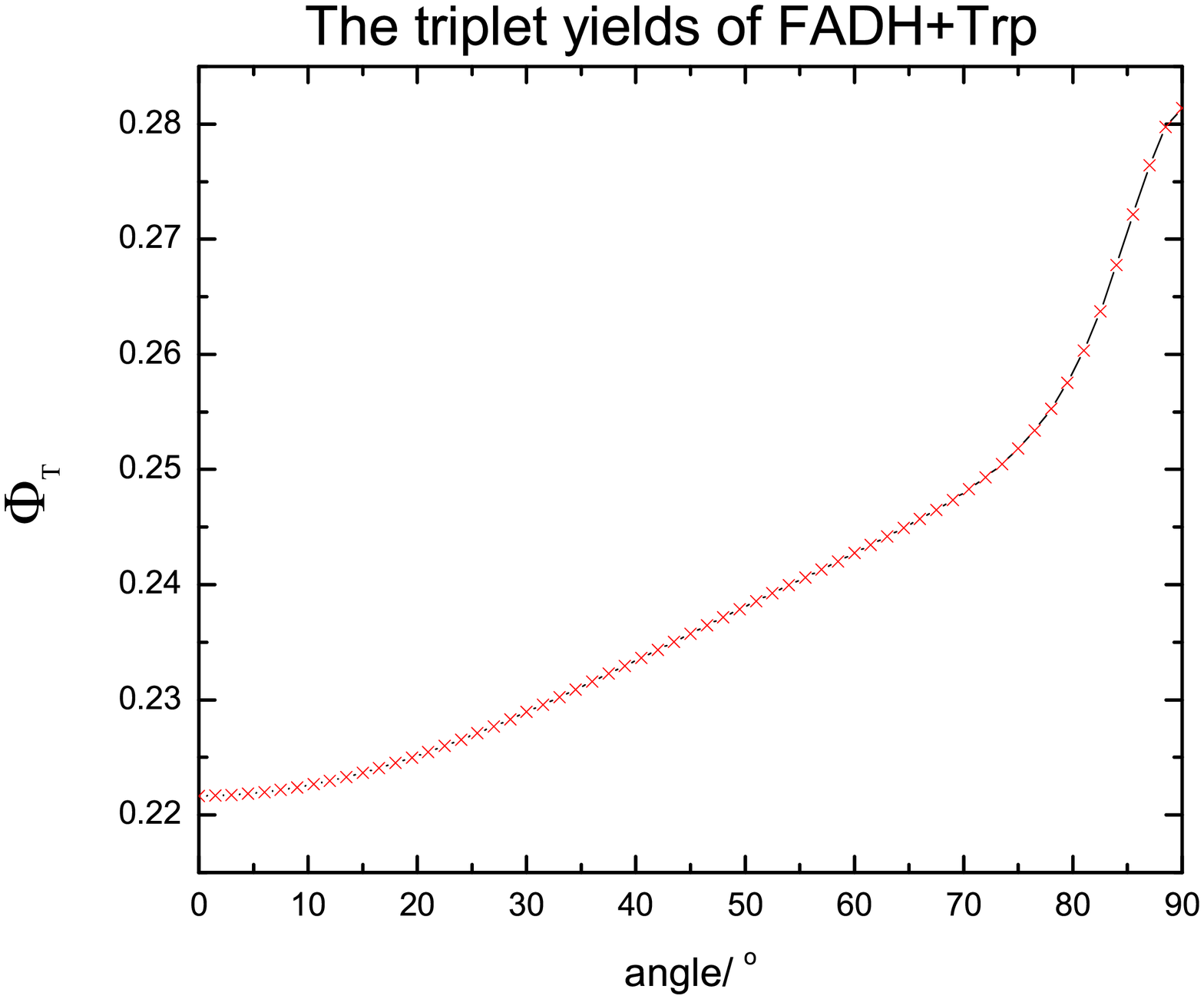}
	\includegraphics[width=0.45\textwidth]{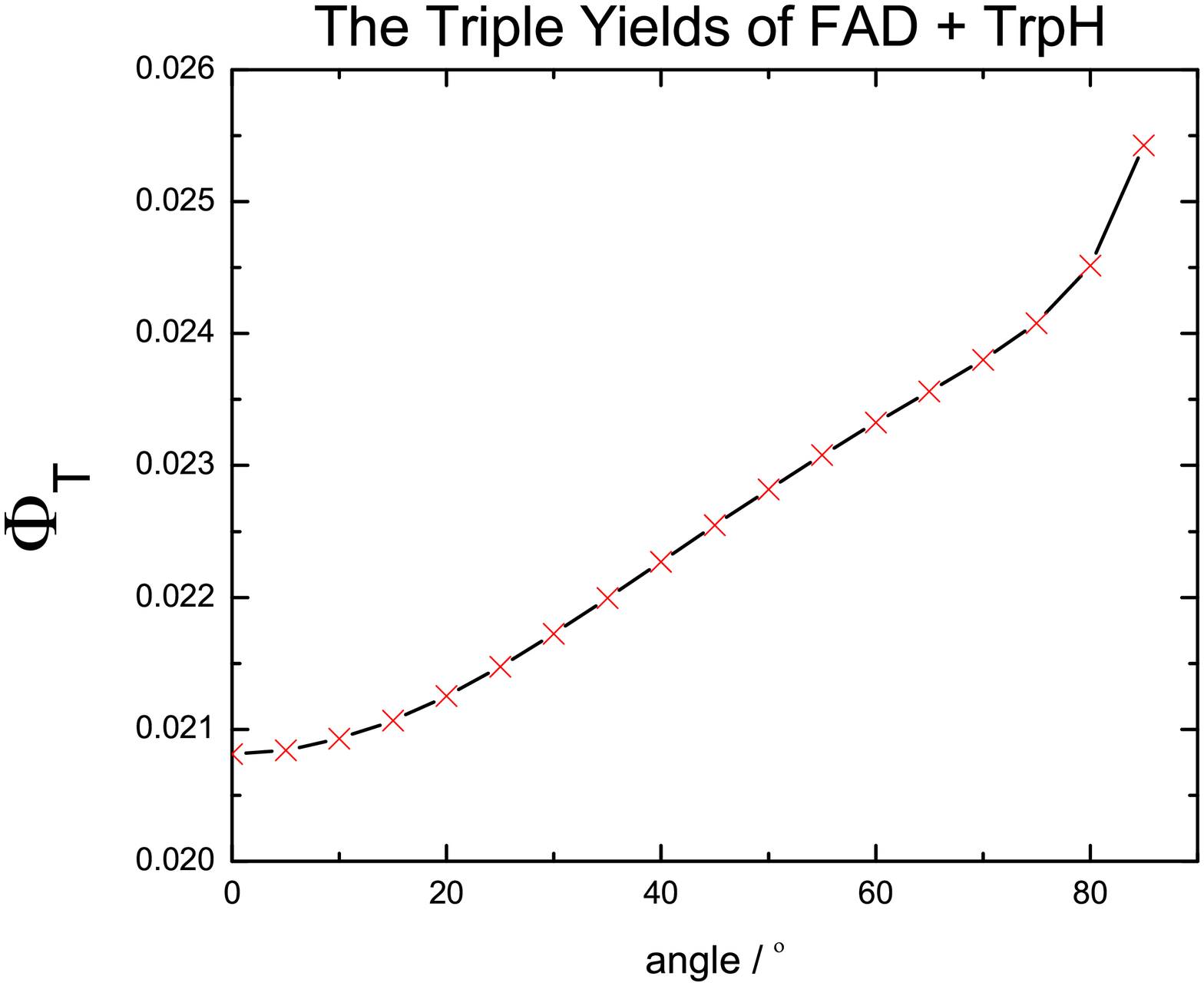}
	\caption{The triplet yields of radical pairs [FADH$^\bullet$Trp$^\bullet$] and [FAD$^\bullet$$^{-}$TrpH$^\bullet$$^{+}$] as a function of angles. The figures show that the yield of the pair [FAD$^\bullet$$^{-}$TrpH$^\bullet$$^{+}$] is almost zero. That means, after a relative long time, the radical pair [FAD$^\bullet$$^{-}$TrpH$^\bullet$$^{+}$] has converted to the pair [FADH$^\bullet$Trp$^\bullet$] via chemical reactions. Thus, the triplet yield of [FAD$^\bullet$$^{-}$TrpH$^\bullet$$^{+}$] is almost zero.}
	\label{fig2}
\end{figure}
\\
Fig. \ref{fig2} shows the yields of [FAD$^\bullet$$^{-}$TrpH$^\bullet$$^{+}$] and [FADH$^\bullet$Trp$^\bullet$]. We can tell that after a relatively long time, the triplet yield of [FAD$^\bullet$$^{-}$TrpH$^\bullet$$^{+}$] is almost zero, which demonstrates that the pair of [FAD$^\bullet$$^{-}$TrpH$^\bullet$$^{+}$] has transferred to [FADH$^\bullet$Trp$^\bullet$] via some chemical reactions. Basically, following the scheme we used, these results of yields can be important, and the yield of [FADH$^\bullet$Trp$^\bullet$] can be seen as the signal for birds. Around 90$^\circ$ (80$^\circ$-90$^\circ$), the derivative of yields with respect to the angle seems to be larger than for the other angles, which indicates that the birds are more sensitive when they are heading north. This could be a good sign, because it may give birds the cue of direction.
\begin{figure}[htpb]
\begin{center}
\includegraphics[width=0.8\textwidth]{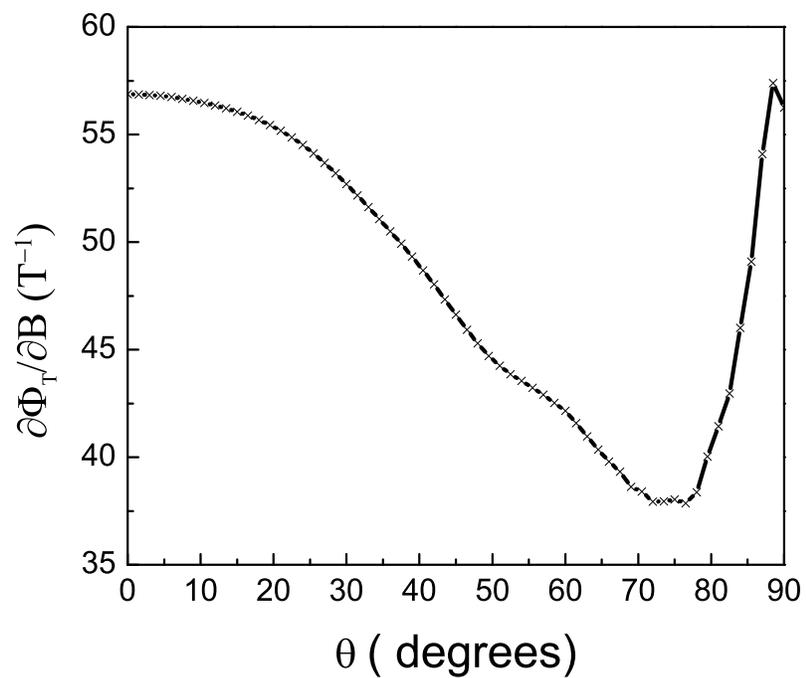}
\end{center}
\caption{\small{The magnetic sensitivity of the chemical compass as a function of the angle. The sensitivity is defined as $\partial \Phi_T/\partial B $, in T$^{-1}$. There is a rapid increase in this sensitivity between 80$^{\circ}$ and 90$^{\circ}$. The sensitivity under geomagnetic field is of the order of $10^{-3}$, requiring a strong magnification mechanism in birds to utilize it.}}
\label{senang}
\end{figure}
\\
We now focus on the magnetic sensitivity of the avian compass, which is defined as $\partial \Phi_T/\partial B$ (T$^{-1}$) \cite{et4, drrp}. Fig. \ref{senang} shows that the sensitivities around 0$^{\circ}$ and 90$^{\circ}$ are similar and also larger than for most other angles, which could indicate that the birds can detect the directions of meridians and parallels if they use the intensity of the magnetic field for navigation, since the yield-based compass is most sensitive along these two directions. Another property that attracted our attention is that the sensitivity's slope is significantly larger between 80$^{\circ}$ and 90$^{\circ}$ than that of the other sections of the curve. This property of increased sensitivity may imply that it is easier for birds to detect the direction of magnetic parallels than that of magnetic meridians. Since the yield-based compass is very sensitive to the change of intensities, we can also expect that it is easier for birds to detect the change of the field intensities when the polar angle is around 90$^{\circ}$. This capability can enable birds to migrate along the direction of the gradient of intensities of the geomagnetic fields.
\begin{figure}[htpb]
\begin{center}
\includegraphics[width=0.8\textwidth]{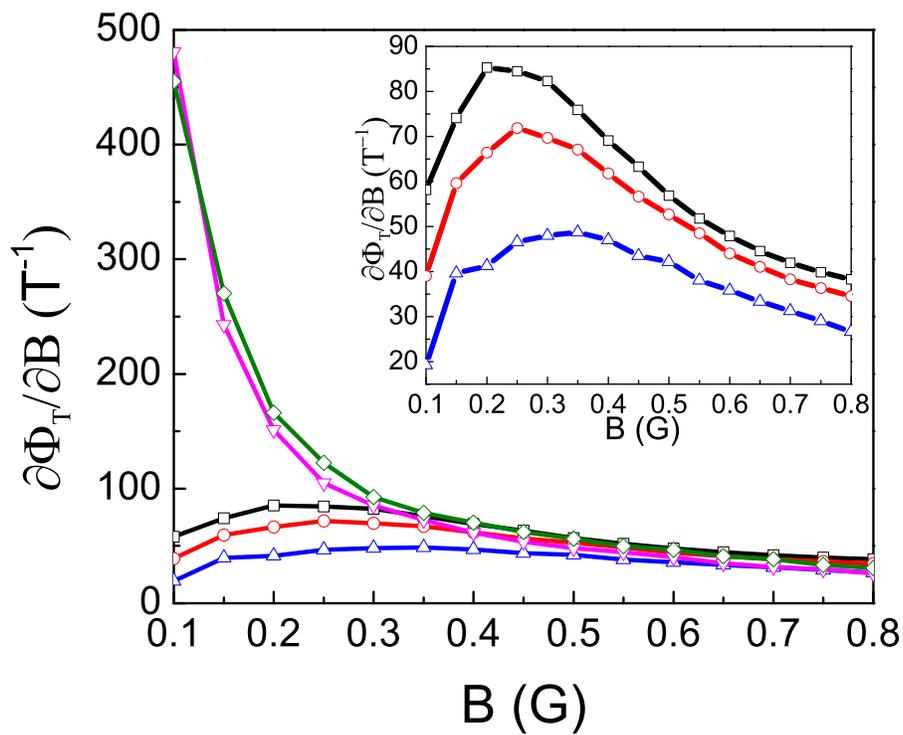}
\end{center}
\caption{\small{(Color online) Intensity dependence of the magnetic sensitivity. This graph shows the sensitivity as a function of external magnetic field, $\vec{B}$, as a function of polar angle, $\theta$, between the z-axis of the radical pair and the magnetic field, i.e., $\theta = 0^{\circ}(black)$, $30^{\circ}(red)$, $60^{\circ}(blue)$, $85^{\circ}(pink)$, $90^{\circ}(green)$. The interior graph magnifies the data for angles $\theta = 0^{\circ}(black)$, $30^{\circ}(red)$, $60^{\circ}(blue)$.}}
\label{sen}
\end{figure}
\\
Furthermore, we explore the magnetic sensitivity as a function of the intensities of the magnetic field for several polar angles $\theta$, in Fig. \ref{sen}. Fig. \ref{sen} shows two types of curves. The first pattern is observed for 85$^{\circ}$ and 90$^{\circ}$, in which the sensitivities monotonically decrease as the external fields increase. In this situation, the sensitivities are much higher for very weak magnetic fields, less than 0.25G, than those in the normal range of the geomagnetic fields, from 0.25G to 0.65G \cite{WMM}. The sensitivities fall into the normal range in the geomagnetic fields, similar to other angles. The other pattern occurs for 0$^{\circ}$, 30$^{\circ}$ and 60$^{\circ}$, and the sensitivities increase initially, and then decrease as the external fields increase. In this situation, the maxima of the curves move rightwards and downwards as the polar angles increase. Combining these two situations (Fig. \ref{sen}), we observe the properties of the chemical compass mentioned before, namely that compass is most magnetically sensitive around 0$^{\circ}$ and 90$^{\circ}$ at the geomagnetic fields. However, above 0.35G, all sensitivities decrease as the fields' intensities increase. This may explain why some species of birds lose their ability to orient themselves in higher magnetic fields \cite{cc0, ww1}. Also, since the sensitivity is not zero, after extended exposure to unnatural magnetic fields the birds may adapt to the decreased sensitivity, so that they are able to regain the ability to orient \cite{regain}. 
\\
\\
All of the above results can provide us with a basic picture of the radical pair mechanism.
\section{Conclusions}
The radical pair mechanism is a promising hypothesis to explain the mystery of the navigation of birds. This theoretical study has demonstrated the role of weak magnetic fields play in the product yields of the radical pairs. In addition, this type of study has inspired scientists to design highly effective devices to detect weak magnetic fields and to use the geomagnetic fields to navigate.
\\
\\
The anisotropic hyperfine coupling between the electron spins and the surrounding nuclear spins can play a crucial role in avian magnetoreception. The hyperfine coupling can affect not only the product yields but also the entanglement of the electron spin states. By involving more nuclear spins one can greatly enhance the quantum entanglement \cite{nuc}. Additionally, mimicking this anisotropic magnetic environment can be very useful for creating detectors of weak magnetic fields.
\\
\\
By studying the role of intensity of the magnetic field in avian navigation, we find that birds could be able to detect the change of the intensity of geomagnetic fields and the approximate direction of parallels instead of sensing the exact direction. However, the mechanism in which birds can utilize the signal remains unknown at this time.

\subsection*{Acknowledgements} S.K. and Y.Z. would like to thank the NSF Center for Quantum Information for Quantum Chemistry (QIQC), Award
No. CHE-1037992, for financial support. The work by G.P.B. was carried out under the auspices of the National Nuclear Security Administration of the U.S. Department of Energy at Los Alamos National Laboratory under
Contract No. DE-AC52-06NA25396.
 

\end{document}